\documentclass{appolb}
\usepackage{epsfig}
\usepackage{cite}

\begin{document}

\eqsec  

\title{PENTAGON DIAGRAMS OF BHABHA SCATTERING%
\thanks{Presented by K.~Kajda.\, 
Work supported in part by
Sonderforschungs\-be\-reich/Transregio TRR 9 of DFG
`Computerge\-st{\"u}tz\-te Theo\-re\-ti\-sche Teil\-chen\-phy\-sik',
and by
the European Community's Marie-Curie Research Training Networks
MRTN-CT-2006-035505 `HEPTOOLS' and MRTN-CT-2006-035482 `FLAVIAnet'.}
}
\author{
J. Fleischer$^{a,b}$, J. Gluza$^c$, K. Kajda$^c$, T. Riemann$^b$
\address{
$^a$Fakult\"at f\"ur Physik, Universit\"at Bielefeld\\
Universit\"atsstr. 25, 33615 Bielefeld, Germany
\\[.3cm]
$^b$Deutsches Elektronen-Synchrotron DESY\\
Platanenalle 6, D-15738 Zeuthen, Germany
\\[.3cm]
$^c$Institute of Physics, University of Silesia\\
Uniwersytecka 4, 40-007 Katowice, Poland
}
}
\maketitle

\begin{abstract}
We report on tensor reduction of five point integrals needed for the 
evaluation of loop-by-loop corrections to Bhabha scattering. 
As an example we demonstrate the calculation of the rank two tensor integral with 
cancellation of the spurious Gram determinant 
in the denominator. 
The reduction scheme is worked out for arbitrary
five point processes.
\end{abstract}
\PACS{13.20.Ds, 13.66.De}
\section{Introduction}
Evaluation of the Bhabha cross-section with two-loop accuracy 
\cite{nf1,Penin:2005kf,Actis:2007gi,Becher:2007cu} would be incomplete without taking into account 
contributions from one loop diagrams with photon emission from internal electron lines. 
So far bremsstrahlung from external legs (lowest order) has been taken into account
 \cite{Melles:1996qa} and implemented in \cite{Jadach:1995hy,Jadach:2001jx,Arbuzov:2004wp}.\\

\begin{figure}
  \begin{center}
    \epsfig{file=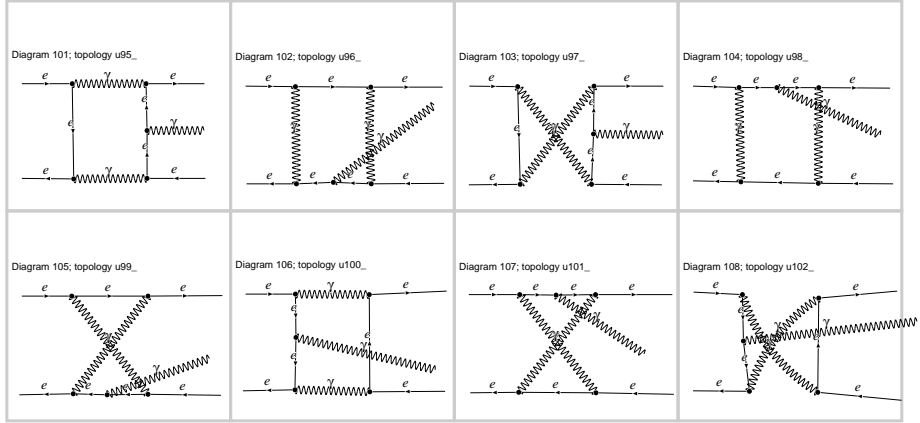,width=12cm}
  \end{center}
  \caption{\label{TreeLoop}
Eight diagrams at one loop level obtained using DIANA \cite{Tentyukov:2002ig}.}
\end{figure}

The general form of the five point tensor integrals in the diagrams of Fig. \ref{TreeLoop} is: 
\begin{equation}
	I_5^{ \{ 1,q^{\mu},q^{\mu}q^{\nu} \dots \} }=
	e^{\epsilon \gamma_E}
	\int \frac{d^d q}{i \pi^{d/2}} 
	\frac{ \{ 1,q^{\mu},q^{\mu}q^{\nu} \dots \} }
	     {c_1 c_2 c_3 c_4 c_5},
\end{equation}
where $c_i=(q+q_i)^2-m^2_i, \quad i=1, \dots ,5$. We also make the (arbitrary) choice $q_5=0$.
~\\
Massive five point functions tend to be unstable numerically in certain
kinematic domains due to the appearance of an inverse Gram
determinant. Calculating the tensor
integrals directly, e.g. by using Mellin-Barnes (MB-) representations,
we have to cope with five dimensional MB-integrals (before expansion in
epsilon). After $\epsilon$ expansion, five dimensional MB
representation still remains at $\epsilon^0$ (e.g. for second order tensor).
The MB-representations are derived with the aid of the Mathematica
packages {\ttfamily AMBRE} and {\ttfamily MB}
\cite{Gluza:2007rt,Czakon:2005rk}.
They are hard to evaluate both numerically (and in fact presently the {\ttfamily MB}
package works in Euclidean region only) and analytically.
A reduction of tensor five point functions to scalar four and three
point functions solves that problem.
\section{Reduction of five point functions}
We give examples of reduction for the scalar and second rank tensor five point function where the spurious Gram determinant 
in denominators is avoided (see also \cite{Fleischer:2007ff}).

To make a reduction one can follow the standard Passarino-Veltman \cite{Passarino:1978jh} reduction scheme, or,  as we will do, follow a scheme based on \cite{Melrose:1965kb,Fleischer:1999hq,Denner}.  
For a diagram with internal lines $1 \dots n$ we can introduce the so called "Modified Cayley Determinant":
\begin{equation}
        ()_5 =
        \left| 
	\begin{array}{cccccc}
        0 & 1       & 1       &\ldots & 1      \\
	1 & Y_{11}  & Y_{12}  &\ldots & Y_{1n} \\
	1 & Y_{12}  & Y_{22}  &\ldots & Y_{2n} \\
	\vdots  & \vdots  & \vdots  &\ddots & \vdots \\
	1 & Y_{1n}  & Y_{2n}  &\ldots & Y_{nn}
        \end{array} \right|,
\end{equation}
with $Y_{ij}=-(q_i-q_j)^2+m_i^2+m_j^2$.
By cutting from $()_5$ rows $j_1, j_2, \dots$ and columns $k_1, k_2, \dots$ we get so-called "Minors".
The sign of a "Signed Minor",
\begin{equation}
        \left( \begin{array}{ccc}
        j_1 & j_2 & \dots \\
        k_1 & k_2 & \dots
        \end{array} \right)_n,
\end{equation}
is determined by the sum of indices of excluded rows and columns and by taking into account the appropriate signatures of the permutations, taken separately from excluded rows and columns (important when cancelling more than one row and one column), e.g. signatures of permutations for ${1 2 \choose 5 4}$ is: $+$ for rows and $-$ for columns. In this way:
\begin{equation}
        {0 \choose 0}_n=
        \left| \begin{array}{ccccc}
        Y_{11}  & Y_{12}  &\ldots & Y_{1n} \\
	Y_{12}  & Y_{22}  &\ldots & Y_{2n} \\
	\vdots  & \vdots  &\ddots & \vdots \\
	Y_{1n}  & Y_{2n}  &\ldots & Y_{nn}
        \end{array} \right|.
\end{equation}
Using "signed minors" one can write tensor reduction formulas in a compact and elegant way.\\ 

We start from the simplest case of a scalar five point function. As starting point we use the recursion relation \cite{Fleischer:1999hq}:
\begin{equation}
          (d-\sum_{i=1}^{n}\nu_i+1)G_{n-1}I^{(d+2)}_n=
          \left[2 \Delta_n+\sum_{k=1}^n 
		\left(
			\frac{\partial \Delta_n} {\partial m_k^2}
		\right) 
	  {\bf k^-} \right]I^{(d)}_n,
\end{equation}
where
\begin{equation}
        \Delta_n={0 \choose 0}_n, \quad
        G_{n-1}=-\sum_{k=1}^{n}
        \left( \frac{\partial \Delta_n}
                    {\partial m_k^2}
        \right) =2()_n.
\end{equation}
The operator {$\bf k^-$} 
decreases the $k$-th propagator power by one. Because in our case powers of propagators are all equal to one,
 and we are interested in the case with $n=5$, the above relation changes into:
\begin{equation}
        (d-4)()_5 I_5^{(d+2)}=
        {0 \choose 0}_5 I_5-
        \sum_{k=1}^{5} {0 \choose k}_5 I_4^{k}.
\end{equation}
In the limit $d \rightarrow 4$ we get as final reduction formula for the scalar five point function:
\begin{equation}
	I_5=\frac{1}{ {0 \choose 0}_5 } \sum_{k=1}^5 {0 \choose k}_5 I_4^k,
\end{equation}
where $I_4^k$ is the scalar four point integral obtained by cancelling the $k$-th line in the 
five point function diagram.\\
For the vector integral we refer to \cite{Fleischer:1999hq}. The second rank tensor integral can be written in the form:
\begin{equation}
	I_{5}^{\mu\, \nu} = 
	\sum_{i,j=1}^{4} q_i^{\mu}q_j^{\nu}
	n_{ij} I_{5,ij}^{[d+]^2} -
	\frac{1}{2} g^{\mu \nu} I_{5}^{[d+]}.
\end{equation}
Here $[d+]^l=4+2l-2 \epsilon$, and $n_{ij}=1+\delta_{ij}$. We start from the following formulas \cite{Fleischer:1999hq}:
\begin{equation}
	n_{ij} I_{5,ij}^{[d+]^2}=
	-\frac{{0\choose j}_5}{\left(  \right)_5} 
	I_{5,i}^{[d+]} + \sum_{s=1,s \ne i}^{5} 
	\frac{{s\choose j}_5}{\left(  \right)_5} 
	I_{4,i}^{[d+],s} +
 	\frac{{i\choose j}_5}{\left(  \right)_5} 
	I_{5}^{[d+]},
 \label{I5ij}
\end{equation}
\begin{equation}
	g^{\mu\, \nu}=
	2 \sum_{i,j=1}^{4} 
	\frac{{i\choose j}_5}{\left(  \right)_5} 
	q_i^{\mu}q_j^{\nu}, 
 \label{gmunu}
\end{equation}
\begin{equation}
	I_{4,i}^{[d+],s}=
	-\frac{{0s\choose is}_5}{{s\choose s}_5} 
	I_{4}^{s} +
 	\sum_{t=1,t \ne s}^{5} 
	\frac{{ts\choose is}_5}{{s\choose s}_5} 
	I_{3}^{st}.
\end{equation}
Using all the above contributions we end up with:
\begin{equation}
	I_5^{\mu \nu}=
	\sum_{i,j=1}^{4} 
	q_i^{\mu}q_j^{\nu} I_{5,ij},
\end{equation}
where	
\begin{equation}
  I_{5,ij}=
  \frac{1}{\left(  \right)_5} 
  \left\{-\frac{{0\choose j}_5}{{0\choose 0}_5}  
	\sum_{s=1}^{5} {0i\choose 0s}_5 I_{4}^{s} -
	\sum_{s=1}^{5} 
  \frac{{s\choose j}_5 {0s\choose is}_5}{{s\choose s}_5} I_{4}^{s} +
  	\sum_{s,t=1}^{5} 
  \frac{{s\choose j}_5 {ts\choose is}_5}{{s\choose s}_5} I_{3}^{st}\right\}.
\label{I5ija}
\end{equation}
However, we can avoid the determinant $()_5$ in the denominator. We begin with the following structure:
\begin{equation}
	I_5^{\mu \nu}=
	[ I_5^{\mu \nu}-
	E_{00}g^{\mu \nu} ]+
	E_{00}g^{\mu \nu}
\label{ansatz}
\end{equation}
The problem of finding an appropriate ansatz for $E_{00}$ has been solved in \cite{Denner}:
\begin{equation}
	E_{00}=
	\frac{1}{ {0 \choose 0} }
	\sum_{s=1}^{5}
	{0 \choose s }_5 D_{00}^s ,
\end{equation}
where $D_{00}^s$ is the $g^{\mu \nu}$ term of the tensor four point function. The complete result for $E_{00}$ is:
\begin{equation}
	E_{00}=
       -\frac{1}{2} \frac{1}{ {0 \choose 0}_5 }	
	\sum_{s=1}^{5} \frac{ {0 \choose s}_5 }{ {s \choose s}_5 }	
	\Bigg[
		{0s \choose 0s}_5 I_4^s-\sum_{t=1}^5 {0s \choose ts}_5 I_3^{st}
	\Bigg].
  \label{E00}
\end{equation}
To demonstrate in details how $()_5$ is cancelled in the square bracket in (\ref{ansatz}), we have to consider four and three point functions separately by analyzing the coefficients of $I_4^s$ and  $I_3^{st}$
in (\ref{I5ij}), subtracting the $g^{\mu \nu}$ term according to (\ref{gmunu}) and (\ref{E00}). For
$I_4^s$ we have
\begin{eqnarray}
&& \frac{1}{{\left(  \right)_5}{0\choose 0}_5 {s\choose s}_5} 
	\times
\nonumber\\
&& \left\{- {0\choose j}_5 {0s\choose 0i}_5 {s\choose s}_5 -
{s\choose j}_5 {0s\choose is}_5 {0\choose 0}_5 + {0\choose s}_5  {0s\choose 0s}_5 {i\choose j}_5 \right\}_{ij}^s
\nonumber\\
&& \equiv \frac{1}{{0\choose 0}_5 {s\choose s}_5} X_{ij}^s, 
\label{cancel4}
\end{eqnarray}
i.e. we have to show that indeed $()_5$ cancels and we have to give an explicit expression for $X_{ij}^s$.\\

A useful property of
  $X_{ij}^s$ is its symmetry w.r.t. the indices $i$ and $j$ for fixed $s$. Obviously the third term in the curly bracket of (\ref{cancel4}) is symmetric since we consider a symmetric determinant. The symmetry of the first two terms means
\begin{eqnarray}
	&& {s\choose s}_5 \left[{0\choose i}_5 
	   {0j\choose 0s}_5-{0\choose j}_5 
	   {0i\choose 0s}_5 \right] 
\nonumber\\
	&+& {0\choose 0}_5 \left[{s\choose i}_5 
	    {0s\choose js}_5-{s\choose j}_5 
	    {0s\choose is}_5 \right]=0.
\label{vanish}
\end{eqnarray}
The first square bracket of (\ref{vanish}) can be evaluated using (A.13) of \cite{Melrose:1965kb}, i.e.
\begin{equation}
	{0\choose j}_5 {0i\choose 0s}_5=
	-{0\choose 0}_5 {0s\choose ij}_5+
	{0\choose i}_5 {0j\choose 0s}_5
\end{equation}
and (\ref{vanish}) then results in:
\begin{equation}
	{s\choose i}_5 {0s\choose js}_5+
	{s\choose j}_5 {0s\choose si}_5+
	{s\choose s}_5 {0s\choose ij}_5=0.
\label{circle}
\end{equation}
This is proved by multiplication with $()_5$ and using (A.8) of \cite{Melrose:1965kb} with $r=2$,~ i.e.
\begin{equation}
{il\choose jk}_5 {\left(  \right)_5} = {i\choose j}_5 {l\choose k}_5 - {i\choose k}_5 {l\choose j}_5.
\label{r2}
\end{equation}
Inserting this, products of three factors of the form ${i\choose k}_5$ cancel pairwise, q.e.d.~.\\

Further, the following relations (A.11) and (A.12) of \cite{Melrose:1965kb} are important, i.e.
\begin{equation}
	\sum_{i=1}^n {0\choose i}_5 =()_5,
	\qquad
	\sum_{i=1}^n {j\choose i}_5 =0,~ (j \ne 0).
  \label{A11A12}
\end{equation}
As simplest case we now immediately obtain from (\ref{cancel4}) $\{ \cdots \}_{ss}^s=0$, i.e. $X_{ss}^s=0$.
Applying  (\ref{A11A12}) to  (\ref{cancel4}), we see:
\begin{equation}
	\sum_{j=1}^5 \{ \cdots \}_{ij}^s = 
	-{\left(  \right)_5} 
	 {0s \choose 0i}_5  
	 {s \choose s}_5
 \label{sumj}
\end{equation}
and due to the symmetry in $i$ and $j$ we also have:
\begin{equation}
	\sum_{i=1}^5 \{ \cdots \}_{ij}^s = 
	-{\left(  \right)_5} 
	{0s\choose 0j}_5  
	{s\choose s}_5,
 \label{sumi}
\end{equation}
which gives us a hint of how $X_{ij}^s$ might look, namely due to  (\ref{sumj}) it should contain a term $-{0s\choose 0i}_5 {0s\choose js}_5$. A further contribution, summed over, must vanish. Due to $X_{ss}^s=0$ it must contain a factor $ {0j\choose si}_5$. The second factor of this further contribution can only depend on $s$ and has been determined by explicit calculation to be ${0s\choose 0s}_5$. Thus we conclude
\begin{equation}
	X_{ji}^s=X_{ij}^s=
	-{0s\choose 0i}_5 {0s\choose js}_5+ 
	 {0j\choose si}_5 {0s\choose 0s}_5.
 \label{Xijs}
\end{equation}
~\\

For $I_3^{st}$ a slight generalisation yields the coefficient
\begin{equation}
	X_{ij}^{st}=
	-{0s\choose 0j}_5 {ts\choose is}_5+ 
	 {0i\choose sj}_5 {ts\choose 0s}_5,
 \label{Xijst}
\end{equation}
with $X_{ij}^{s}=X_{ij}^{s,t=0}$ and the final result is: 
\begin{equation}
	I_5^{\mu \nu}=
	\sum_{i,j=1}^{4} 
	q_i^{\mu}q_j^{\nu} I_{5,ij}+
	g^{\mu \nu} E_{00},
\end{equation}
where	
\begin{equation}
	I_{5,ij}=
	\frac{1}{ {0 \choose 0}_5 }
	\sum_{s=1}^{5} \frac{1}{ {s \choose s}_5 }
	\Big(
		X_{ij}^{s0}I_4^s-\sum_{t=1}^5 X_{ij}^{st}I_3^{st}
	\Big).	
\end{equation}
$X_{ij}^{st}$ is defined in (\ref{Xijst}) and $E_{00}$ in \ref{E00}. $I_3^{st}$ is the scalar three point function obtained by cutting the $s$ and $t$ lines in the five point diagram.

\section{Automatization}
We have developed a Mathematica package which reduces general 
tensor five point functions up to rank three. 
Using {\ttfamily LoopTools} \cite{Hahn:1998yk} we have performed numerical cross-checks which ensured us about the correctness of the package. 
We have also made independent checks with sector decomposition and  the Mellin-Barnes method. 
Some examples are given in \cite{cross-checks}.

In summary, we have developed tools to deal with five point Bhabha amplitudes. 
The next step is to implement it into the complete numerical 
calculation.



\begin{thebibliography}{99}

\bibitem{nf1}
  R.~Bonciani, A.~Ferroglia, P.~Mastrolia, E.~Remiddi and J.~J.~van der Bij,
  Nucl.\ Phys.\  B {\bf 701} (2004) 121
  [arXiv:hep-ph/0405275];\\
  R.~Bonciani, A.~Ferroglia, P.~Mastrolia, E.~Remiddi and J.~J.~van der Bij,
  Nucl.\ Phys.\  B {\bf 716} (2005) 280
  [arXiv:hep-ph/0411321].

\bibitem{Penin:2005kf}
  A.~A.~Penin,
  Phys.\ Rev.\ Lett.\  {\bf 95} (2005) 010408
  [arXiv:hep-ph/0501120].

\bibitem{Actis:2007gi}
  S.~Actis, M.~Czakon, J.~Gluza and T.~Riemann,
  Nucl.\ Phys.\  B {\bf 786} (2007) 26.
  [arXiv:0704.2400 [hep-ph]];\\
T.~Riemann, ``Fermionic NNLO corrections to Bhabha scattering'', talk held at this conference and arXiv:0710.5111 [hep-ph]


\bibitem{Becher:2007cu}
  T.~Becher and K.~Melnikov,
  JHEP {\bf 0706} (2007) 084
  [arXiv:0704.3582 [hep-ph]].

\bibitem{Melles:1996qa}
  M.~Melles,
  Acta Phys.\ Polon.\  B {\bf 28} (1997) 1159
  [arXiv:hep-ph/9612348].


\bibitem{Jadach:1995hy}
  S.~Jadach, M.~Melles, B.~F.~L.~Ward and S.~A.~Yost,
  Phys.\ Lett.\  B {\bf 377} (1996) 168
  [arXiv:hep-ph/9603248].

\bibitem{Jadach:2001jx}
  S.~Jadach, M.~Melles, B.~F.~L.~Ward and S.~A.~Yost,
  Phys.\ Rev.\  D {\bf 65} (2002) 073030
  [arXiv:hep-ph/0109279].

\bibitem{Arbuzov:2004wp}
  A.~B.~Arbuzov, D.~Haidt, C.~Matteuzzi, M.~Paganoni and L.~Trentadue,
  Eur.\ Phys.\ J.\  C {\bf 34} (2004) 267
  [arXiv:hep-ph/0402211].

\bibitem{Gluza:2007rt}
  J.~Gluza, K.~Kajda and T.~Riemann, to appear in Comput.\ Phys.\ Commun.
  [arXiv:0704.2423 [hep-ph]].

\bibitem{Czakon:2005rk}
  M.~Czakon,
  Comput.\ Phys.\ Commun.\  {\bf 175} (2006) 559
  [arXiv:hep-ph/0511200].

\bibitem{Tentyukov:2002ig}
  M.~Tentyukov and J.~Fleischer,
  Nucl.\ Instrum.\ Meth.\  A {\bf 502} (2003) 570
  [arXiv:hep-ph/0210179].

\bibitem{Fleischer:2007ff}
J.~Fleischer,
{\em  Application of Mellin-Barnes representation to the
 calculation of massive five-point functions in Bhabha scattering},
 talk given at the Conference on Frontiers in Perturbative Quantum Field
 Theory, June 14-16 2007, ZiF, Bielefeld.

\bibitem{Passarino:1978jh}
  G.~Passarino and M.~J.~G.~Veltman,
  Nucl.\ Phys.\  B {\bf 160} (1979) 151.

\bibitem{Melrose:1965kb}
  D.~B.~Melrose,
  Nuovo Cim.\  {\bf 40} (1965) 181.

\bibitem{Fleischer:1999hq}
  J.~Fleischer, F.~Jegerlehner and O.~V.~Tarasov,
  Nucl.\ Phys.\  B {\bf 566} (2000) 423
  [arXiv:hep-ph/9907327].

\bibitem{Denner}
  A.~Denner and S.~Dittmaier,
  Nucl.\ Phys.\  B {\bf 658} (2003) 175
  [arXiv:hep-ph/0212259];\\
  A.~Denner and S.~Dittmaier,
  Nucl.\ Phys.\  B {\bf 734} (2006) 62
  [arXiv:hep-ph/0509141].

\bibitem{Hahn:1998yk}
  T.~Hahn and M.~Perez-Victoria,
  Comput.\ Phys.\ Commun.\  {\bf 118} (1999) 153
  [arXiv:hep-ph/9807565].

\bibitem{cross-checks}
Transparencies of this talk, http://www.us.edu.pl/$\sim$us2007.
\end{thebibliography}
\end{document}